\documentclass[12pt]{article}
\usepackage{amsmath, latexsym, amssymb}

\numberwithin{equation}{section}
\newcommand{\R}{\mathbb R}
\newcommand{\C}{\mathbb C}

\newcommand{\X}{\mathbb{X}}

\newcommand{\w}{\bar{w}}

\newcommand{\Ta}{T_0}

\newcommand{\T}{T_0'}

\newcommand{\Bold}[1]{{\boldsymbol{\mathit{#1}}}}
\newcommand{\dual}{\langle\tau',\tau\rangle}
\newcommand{\zdual}{\langle b',b\rangle}

\newcommand{\ti}{\mathrm{t}}

\newcommand{\x}{\mathrm{x}}
\newcommand{\s}{\mathrm{s}}

\newcommand{\xx}{\bar{x}}

\newcommand{\M}{\mathbb{M}}

\newcommand{\z}{b}

\newcommand{\QED}{\hspace{.2in}\square\newline}

\newtheorem{proposition}{Proposition}[section]
\newtheorem{definition}{Definition}[section]

\begin{document}

\begin{center}
{\Large \textbf{Functional Integration on\\Constrained Function
Spaces I: Foundations}} \vskip 2em
{ J. LaChapelle}\\
\vskip 2em
\end{center}

\begin{abstract}
 Analogy with Bayesian inference is used to formulate constraints within a scheme for functional integration proposed by Cartier and DeWitt-Morette. According to the analogy, functional counterparts of
conditional and conjugate probability distributions are introduced
for integrators. The analysis leads to some new
functional integration tools and methods that can be applied to the study of constrained dynamical systems.
\end{abstract}

\vskip 1em


\vskip 1em



\section{Introduction}
Kinematical constraints (e.g. constraints in the form of boundary conditions) on dynamical systems
modeled by differential equations have been well-studied. Typical solution
methods are often based on elementary techniques that rely on simple
boundary value matching. But subtleties can arise from complicated geometries/topologies, and it becomes necessary to extend the elementary methods --- especially in quantum
physics. For example, one extension to general
geometries makes use of the generalized Green's theorem: By
formulating the solution of a differential equation in terms of
Green's functions, arbitrary boundary geometries with certain
regularity conditions can be treated. Such extensions
deal directly with function spaces and the mathematical complexities
and subtleties inherent in them.

On the other hand, dynamically constrained systems (e.g. systems with local symmetries) and their
quantization have been --- and continue to be --- extensively
studied for obvious reasons. Solution methods for this constraint
type are usually anything but elementary.  The vast literature on
this topic supports the contention that, here also, the function
spaces of the dynamical variables (as opposed to their target
manifolds) are of primary importance.

Importantly, from a function space perspective, the distinction
between kinematical constraints and dynamical
constraints is unnecessary. Both types can be formulated by posing a
restricted (or constrained) function space: In practice, the restriction is often
imposed indirectly on a target manifold, and it leads to some kind
of set-reduction in some appropriate general function space. Consequently, one can
anticipate that function spaces furnish a fruitful arena in which to
formulate and study \emph{all} constraints.

Moreover, it has long been recognized that functional integration offers
reliable if not always acceptably rigorous methods to probe function
spaces, so it is not surprising that functional integrals have
become useful analytical and numerical tools to study complex,
constrained dynamical systems. Accordingly, they offer a means to
incorporate and study both kinematical and dynamical constraints under one
roof.

There are many references in the physics literature that study
constraints in functional integrals; largely utilizing formal/heuristic or time-slicing approaches.\footnote{Formal/heuristic and time-slicing methods are not without merit: Since functional integrals often localize to Riemann-Stieltjes integrals, the standard manipulations \emph{usually} lead to a correct result.} For a sample, see
\cite{GROS}--\cite{KL}: On the mathematics side, see \cite{SK},\cite{CH} and references therein. The aim of this article is not to supplant those methods
--- they are certainly useful tools --- but to propose a mathematical basis
for functional integration on constrained function spaces. The basis
is suggested by analogy to Bayesian inference theory, and it affords some guiding principles.  With guiding
principles in place, useful integration techniques can be
developed and tested.

This work will utilize the Cartier/DeWitt-Morette (CDM) scheme as
the mathematical foundation of functional integration (without constraints). A short
summary is given in appendix \ref{appendix 1}, but the reader is
encouraged to consult \cite{CA/D-W3}--\cite{CA/D-M} for background
and details. Their approach is similar to, but generalizes, the framework of Albeverio/H{\o}egh-Krohn \cite{AL}. Roughly stated, the CDM scheme uses algebraic duality to define linear integral operators on Banach spaces in terms of bona fide measures on Polish spaces. The as-defined functional integrals, which can be characterized by associated \emph{integrators}, then inherit useful properties through their duality relationship. And these properties can be used to reliably manipulate the functional integrals.\footnote{Since the CDM scheme is restricted to function spaces whose elements are pointed paths that take their values in some
manifold $\X$, i.e. maps $x:[\ti_a,\ti_b]\subset\R\rightarrow\X$
with a fixed point $x(\ti_a)=\x_{a}\in\X$, the functional integrals
in this article are strictly \emph{path integrals}. However, the CDM
scheme can be extended to include more general function spaces (see
e.g. \cite{LA1}) and the guiding principles we identify are not
particular to path integrals in this restricted sense. So the term
functional integral will be used interchangeably with path integral.}

Application of the CDM scheme to unbounded quantum mechanical (QM) systems is well-understood, but how it works under general constraints seems to require new principles.  We start by presenting several well-known examples that contain
clues to the underlying principles. To begin with, they suggest that constraints add non-dynamical degrees of freedom, and this requires an enlarged function space. Next, the Bayesian analogy
suggests the notions of \emph{conditional integrators} and
\emph{conjugate integrators}. Together with the functional counterpart of the Fubini theorem, these tools enable us to construct and  manipulate functional integral representations of constrained dynamical systems within the CDM scheme.

This is the main idea of the paper: Constrained dynamical systems require a state-space comprising dynamical \emph{and} non-dynamical degrees of freedom. Specifying particular constraints induces a subset of the general state space that we designate as a constrained function space. The dynamics of a specific system is then represented by a functional integral based on an appropriate integrator and constrained function space. Finally, the integral over the constrained function space is represented by a functional integral over the full state space but characterized by some conditional integrator.  The conditional integrator is defined by the functional integral analogs of marginal and conditional probability densities, which in turn are related using the Bayesian inference analogy.

The essence of the main idea is just a generalization of the familiar technique often employed to study systems with constraints especially due to symmetry: The physical state/phase space --- which respects the system constraints --- is replaced by a state/phase space that `forgets' the constraints, and the system constraints are implemented through appropriate functionals formulated in terms of some target manifold. The value of encoding constraints within conditional integrators in the CDM scheme lies in a shift of perspective from the target manifold to the function space; which affords a probability interpretation along with its guiding intuitions.

Although the primary focus of the paper is a proposed construction of constrained functional integrals, there are some secondary results obtained along the way that we should point out: 1) The Gaussian integrators in the CDM scheme are redefined to include a boundary form and a parameter that encodes a mean path. The new definitions  are more useful in the context of constraints with their concomitant sufficient statistics. 2) The complex counterpart of the new Gaussian integrator is likewise defined. Although we do not pursue the idea here, it appears that the complex Gaussian might contain important information regarding the Schr\"{o}dinger$\leftrightarrow$diffusion correspondence. Specifically, it might explain when analytic continuation succeeds or fails in this context. 3) In \cite{LA2} a new integrator within the CDM scheme was introduced based on analogy to a gamma probability distribution. Its utility for incorporating boundary conditions in path integrals was recognized, but its meaning and origin were unclear. Here we learn that the gamma integrator is a natural consequence of the Bayesian analogy. Moreover, the gamma integrator possesses a complex parameter that, when restricted to the natural numbers, reduces to what can be characterized as a Poisson integrator. In consequence, the `propagator' for a dynamical system characterized by a gamma integrator yields an equivalent construction of the Poisson functional integral introduced in \cite{CA/D-W3}.

(A caveat; all variables are assumed unit-less by appropriate normalization
for convenience.)

\section{Motivating examples}\label{motivation}
In this section we briefly look at some well-known functional integrals of certain constrained dynamical systems derived using standard semi-heuristic arguments. The exercise is useful as it gives hints about how to do constrained functional integration in general. We will revisit these systems and their constrained functional spaces in more detail in a companion paper \cite{LA3} after developing a firmer mathematical basis in this paper.

\subsection{Localization}
The first class of examples --- constrained Feynman path integrals
--- can be characterized heuristically by the presence of a delta
function(al) in the integrand of a path integral. Some particularly
prevalent early examples of this type in quantum mechanics were
point-to-point transition amplitudes, fixed energy transition
amplitudes, and the propagator for a particle on $S^1$. Let's see what these look like in the CDM scheme for the simplest case of free particles.

In the CDM scheme, the domain of integration for a Gaussian path integral is
a space of pointed paths $X_a$ (see appendix B --- especially for notation). So point-to-point transition
amplitudes are obtained by a suitable delta function in the
integrand that `pins down' the loose end of the paths. Standard
manipulations
\cite{CA/D-W3} reveal that the path integral for quadratic action can
be expressed in terms of a restricted domain of paths
\begin{eqnarray}
\int_{X_{a,b}}e^{2\pi i\langle
x',x\rangle}\mathcal{D}
\omega^{(a,b)}(x)&:=&\int_{X_{a}}\delta\left(x(t_b),x_b\right)e^{2\pi i\langle
x',x\rangle}\mathcal{D}\omega(x)\notag\\
&=& \frac{e^{-\pi i
\mathrm{W}^{(a,b)}(x')}}{\sqrt{\mathrm{det}\left[i\Bold{G}(\ti_b,\ti_b)\right]}}
\end{eqnarray}
where $\Bold{G}(\ti_b,\ti_b)$ is the covariance associated with the
gaussian integrator $\mathcal{D}\omega(x)$ defined on the space of
pointed paths $X_a$ and $\mathrm{W}^{(a,b)}$ is the variance associated with
the \emph{restricted} space of point-to-point paths $X_{a,b}$.

Aside from the action phase factor and the resulting
normalization\footnote{This particular normalization is fixed a
priori from the choice $\int_{X}\mathcal{D}\omega(x)=1$.}
\begin{equation}\label{new normalization}
\int_{X_{a,b}}\mathcal{D} \omega^{(a,b)}(x)=
\frac{1}{\sqrt{\mathrm{det}\left[i\Bold{G}(\ti_b,\ti_b)\right]}}\;,
\end{equation}
the new gaussian integrator $\mathcal{D}\omega^{(a,b)}(x)$, which is defined
on $X_{a,b}$, is characterized by a \emph{different} covariance
$\Bold{G}^{(a,b)}(\ti,\mathrm{s})$ that exhibits the same boundary
conditions as paths in $X_{a,b}$.

Now consider the other two examples. At the classical level,
constraints such as fixed energy and paths on $S^1$ can be imposed
by means of Lagrange multipliers in the classical action. It is then
a standard heuristic argument that the Lagrange multiplier
constitutes a non-dynamical, path-independent degree of freedom in
the path integral that can therefore be integrated out. Essentially, this
introduces what can be characterized as a Dirac delta functional.
However, to give rigorous meaning to a delta functional, one would need a
theory of distributions on $X_a$.

An alternative route is to define a Dirac integrator
$\mathcal{D}\delta(x)$ that does the duty of a delta functional. It
can be thought of as the limit of a Gaussian integrator with vanishing
variance, i.e. $|\mathrm{W}(x')|\rightarrow 0$. The Dirac integrator
reproduces the expected behavior;
\begin{equation}
\int_{X_a} \mathrm{F}(x)\mathcal{D}\delta(x)=\mathrm{F}(0)
\end{equation}
and
\begin{equation}
\int_{X_a} \mathrm{F}(x)\mathcal{D}\delta\left(M(x)\right)
=\sum_{x_0}\frac{\mathrm{F}(x_0)}{\mathrm{Det}M'_{(x_0)}}
\end{equation}
where $M:{X_a}\rightarrow {X_a}$ and $M(x_0)=0$. Otherwise stated, it enforces a
localization in the functional integral domain $X_a$ onto the kernel of $M$.

Similarly, an inverse Dirac integrator $\mathcal{D}\delta^{-1}(x)$
can be formally defined that corresponds to the case
$|\mathrm{W}(x')|\rightarrow \infty$ so that
\begin{equation}
\int_{X_a} e^{-2\pi i \langle x'
,x\rangle}\mathcal{D}\delta^{-1}(x)=\delta(x')\;.
\end{equation}
This integrator enforces a localization in the dual space $X'_a$. In
contrast to a Gaussian or Dirac integrator, this type of integrator however
is not translation invariant;
\begin{eqnarray}
\int_{X_a} e^{-2\pi i \langle x'
,x+x_o\rangle}\mathcal{D}\delta^{-1}(x+x_o) &=&\int_{X_a} e^{-2\pi i
\langle x'+x'_o
,x\rangle}\mathcal{D}\delta^{-1}(x+x_o)=\delta(x')\notag\\
 &\big\Downarrow&\notag\\
 e^{-2\pi i\langle x'_o
,x\rangle}\mathcal{D}\delta^{-1}(x+x_o)
\;&\sim&\;\mathcal{D}\delta^{-1}(x)\;.
\end{eqnarray}
Equivalently,
\begin{equation}
\int_{X_a} e^{-2\pi i \langle x'
,x\rangle}\mathcal{D}\delta^{-1}(x+x_o) =\delta(x'-x'_o)
\end{equation}
where $\langle x'+x'_o,x\rangle:=\langle x',x+x_o\rangle$.

The salient features of note from these three constrained
path integral examples are: i) constraints are related to a
localization in function space (or its dual), ii) constraints are
related to a change in covariance and/or mean, and iii) in general
the normalization of a constrained integrator is different than the
unconstrained integrator.

\subsection{Quotient spaces}
When the target space $\mathbb{X}$ of the pointed paths
$x:[\ti_a,\ti_b]\rightarrow\mathbb{X}$ can be represented as the
base space of a principal fiber bundle $\pi_{\mathbb{G}}:\mathbb{P}{\rightarrow}\mathbb{X}\,$,
\emph{equivariant} forms on $\mathbb{X}$ can be expressed in terms
of associated forms on $\mathbb{P}$. This technique is
essentially symmetry based and allows, for example, determination of propagators
on multiply connected spaces, orbifolds, compact Lie groups, and
homogeneous spaces.

The principal fiber bundle construction is essentially a
generalization of the well-known method of images. In practice, the
group structure of the principal fiber bundle allows the space of paths $X_a$ on the base space to be related
to a group decomposition of the space of paths
$P_a$ on the principal bundle where
$p:[\ti_a,\ti_b]\rightarrow\mathbb{P}$ with $p(\ti_a)=\mathrm{p}_a$. In this way,
paths taking their values in configuration spaces with non-trivial
topology and/or geometry can be treated as paths taking their values
in the covering space. This renders a simplified function space ---
to the extent allowed by the covering space. In terms of path
integrals, the method can be roughly expressed as
\begin{equation}
\int_{X_a}\mathrm{F}(x)\mathcal{D}\omega(x) =
\int_{G}\int_{P_a}\widetilde{\mathrm{F}}(p\cdot g)
\mathcal{D}\widetilde{\omega}(p)\mathcal{D}g \;.
\end{equation}
where $G$ is the space of pointed paths
$g:[\ti_a,\ti_b]\rightarrow\mathbb{G}$ with $g(\ti_a)=\mathrm{g}_a$
and $\mathbb{G}$ the group manifold.

But the functions of interest are equivariant and covariantly
constant which means $\mathrm{p}_a\in\mathbb{P}$ is parallel
transported. This `constraint' induces a map $\mathrm{g}\mapsto
\mathrm{h}\in\mathbb{H}_{(\mathrm{p}_a)}$ into the holonomy group with reference point
$\mathrm{p}_a$, and the integral reduces to the standard result
\begin{equation}
\int_{X_a}\mathrm{F}(x)\mathcal{D}\omega(x)
=\int_{\mathbb{H}_{(\mathrm{p}_a)}}
\int_{P_a}\widetilde{\mathrm{F}}(p\cdot
\mathrm{h})\mathcal{D}\widetilde{\omega}(p)\;d\mathrm{h}\;.
\end{equation}

The point to be made here is that the two function spaces
$P_a$ and $X_a$ are related through a constraint enforced by an integration
(and/or summation for multiply connected or discrete holonomy
groups). Insofar as finite-dimensional integrals are localized
functional integrals, we could loosely say that introducing a
constrained integrator on $P_a\times G$ (implicitly) renders the functional integral on the constrained space ${X_a}$.

\subsection{Discontinuous spaces}
Our final class of examples is comprised of configuration spaces in
which $x(\ti)$ experiences some kind of discontinuity. Particular
cases include bounded configuration spaces, barrier penetration, and tunneling.
The previous two classes of examples gave only a vague hint of how
constraints influence a path integral. However, this third class of
examples yields valuable clues and insights.

If we believe that a variational principle lies at the heart of the
quantum$\rightarrow$classical reduction, then we should re-examine
the variational problem in the context of constraints. Consider a
boundary in configuration space. For point-to-boundary paths, the
correct formulation is a variational problem from a fixed initial
point to a manifold in the dependent-independent variable space.
This type of variational problem introduces a variable end-point
that can be interpreted as a non-dynamical dependent variable that
encodes the implicit constraints imposed by the configuration space
discontinuities and boundaries.

To formulate the variational principle for paths taking their values
in a  space $\X$ that intersects a boundary, consider the
$\mathrm{dim}(n+1)$ dependent--independent variable space
$\mathbb{N}= \X\times\R$ with a terminal manifold of dimension
$(n+1)-k$ defined by some set of equations $\{S_k(\x,\ti)=0\}$ where
$k\leq n$, $\x\in\X$, and $\ti\in [\ti_a,\ti_b]\subset\R$. Let
\begin{equation}
  I(x)=\int_{\ti_a}^{\ti_b}f(\ti,x,\dot{x})\,d\ti\nonumber
\end{equation}
be the functional to be analyzed. The extrema of $I(x)$ solve the
variational problem for point-to-boundary paths. In particular, for
the case of $\X=\R^n$, the variational problem is solved by the
usual Euler equations supplemented by `transversality' conditions
(see e.g. \cite{SA}).

There are two limiting cases of particular interest. When the
terminal manifold in $\mathbb{N}$ coincides with a boundary (or
surface) in $\X$, then $k=1$ and the transversality conditions
reduce to
\begin{equation}\label{transversality}
  f(\ti_b,x(\ti_b),\dot{x}(\ti_b))
  =-\nu\nabla S(x(\ti_b))\cdot\dot{\Bold{x}}(\ti_b)
\end{equation}
where $\nu\neq 0$ is a constant and $x(\ti_b)$ is on the boundary.
For free motion, (\ref{transversality}) implies that critical paths
intersect the boundary transversely.

The other case of interest is when the manifold in the
dependent-independent space is `horizontal', i.e. $x(\ti_b)$ is
fixed and the terminal manifold is a line along the $\ti$ direction.
This clearly corresponds to a point-to-point transition between two
fixed points contained in a bounded region. The terminal manifold is
determined by $k=n$ equations and the transversality conditions
yield
\begin{equation}\label{constant energy}
 f(\ti_b,x(\ti_b),\dot{x}(\ti_b))
  =\nabla_{\dot{\mathbf{e}}} f(\ti_b,x(\ti_b),
  \dot{x}(\ti_b))\cdot\dot{\Bold{x}}(\ti_b)
\end{equation}
where $\dot{\mathbf{e}}$ is a unit vector in the $\dot{\Bold{x}}$
direction. If, in particular, $f=L+E$ where $L$ is the Lagrangian of
an isolated physical system and $E$ is a constant energy, then this
is just the fixed energy constraint $(\partial L/\partial
\dot{x}^i)\dot{x}^i-L=E$. Consequently, the variational problem in
this case is solved by paths with both end-points fixed that have
fixed energy \cite{SA}.

There are two lessons to learn from this: i) when boundaries are
present, we will need to introduce a non-dynamical degree of
freedom, and ii) the boundaries will alter certain expectation
values of the paths according to the transversality conditions.

At this point, the nature of the new degree of freedom is obscure.
However, if one wants a functional integral to represent the
solution of a second order partial differential equation with
non-trivial boundary conditions, then a consistent formulation
emerges if one is willing to associate the new degree of freedom
with a non-Gaussian integrator. It turns out that the new integrator
is closely related to a gamma probability distribution in the same
way that the Gaussian integrator is related to a Gaussian
probability distribution.

The nagging question is, ``Why a gamma integrator?''. The examples
have furnished some clues: not surprisingly they point to
probability theory. If the answer can be understood, perhaps
formulations of path integral representations of more general
differential equations will become evident.

\section{Constraints as conditionals}
Consider a physical system with dynamical, topological, and/or
geometrical constraints. Postulate that such constraints introduce
non-dynamical degrees of freedom. The obvious idea to incorporate these degrees of
freedom in a functional integral context is to enlarge the function space.
Consequently, construct $B\equiv X\times C$ a Banach product space. The space $X$ corresponds to what is usually
thought of as the space of maps, and $C$ will be a space of non-dynamical
degrees of freedom induced by any constraints. In a probability
context, this additional product structure would introduce
conditional and marginal distributions. In our context, we expect analogous structures; about which little can be
said until the nature of the integrators on $C$ are understood.

\subsection{Probability analogy}
Here it is fruitful to develop an analogy with Bayesian inference
theory.\footnote{A rather dated but classic reference for the probability concepts
introduced in this subsection is \cite{RA/SC}; especially Ch. 2.} Momentarily pretend
that $B$ is a probability space. Let $\Theta_X(x)$ and
$\Theta_C(c)$ be the marginal probability distributions on $X$ and
$C$ respectively. Bayes' theorem implies
\begin{eqnarray}\label{Bayes}
\widetilde{\Theta_X}(x|c) &=&\frac{\widetilde{\Theta_C}(c|x)
\Theta_X(x)} {\int_{X}\widetilde{\Theta_C}(c|x) \Theta_X(x)
\,dx}\notag\\\notag\\
&=:&\mathcal{C}(c|x)\Theta_X(x)
\end{eqnarray}
where $\widetilde{\Theta_X}$ and $\widetilde{\Theta_C}$ are
conditional probability distributions on $B$. This yields insight into the constraint
induced normalization noticed in the examples of the previous section.

This induced normalization is not surprising, because a constraint could
alternatively be formulated as a map $M:X\rightarrow Y$ where $y\in
Y$ automatically obeys the constraint. Then change of variable
techniques in the CDM scheme can be used to show the two associated integrators are
related through a functional determinant which is essentially
$\mathcal{C}(c|x)$. This is standard, but it shows that the
probability interpretation is consistent (at least with change of
variables) and it lends credence to the analogy.

So far, we have only made use of Bayes' theorem. To profit further
from the analogy, consider an optical setup where plane
monochromatic waves are focused onto an observation screen. We wish
to study the nature of the light source by placing various
non-conducting apertures between the source and the observation
screen. Sooner or later we discover that under mild intensities the
irradiance pattern on the observation screen is determined by the
mean and covariance of the transmittance at each point in the
aperture. Moreover, by changing the wavelength and/or intensity of
the source, the resulting irradiance pattern can be predicted.

The Bayesian inferential interpretation of these findings is that
the \emph{conditional} probability density or \emph{likelihood}
$\widetilde{\Theta_X}(x|c)$ --- which describes the irradiance
pattern for a given aperture --- can be factored as a product of a
functional $\mathrm{F}(x)$ of the transmittance $x$ and a conditional
likelihood that only depends on the mean and covariance of the
transmittance. In general, the statement is there exist \emph{sufficient
statistics} $S_s(x)$ such that
\begin{equation}\label{conditional theta}
\widetilde{\Theta_X}(x|c)=\mathrm{F}(x)\widetilde{\Theta_{S_s(X)}}(S_s(x)|c)
\end{equation}
where $\mathrm{F}(x)$ is a functional on $X$ and
$\widetilde{\Theta_{S_s(X)}}(S_s(x)|c)$ is a likelihood on
$S_s(X)\times C$. In other words, the irradiance pattern only depends conditionally on a (rather special) subset of $X$. An equivalent statement by way of Bayes' theorem
is that the conditional probability density
$\widetilde{\Theta_C}(c|x)\propto
\Theta_C\left(c\right)\widetilde{\Theta_{S_s(X)}}(S_s(x)|c)$ is a
functional of sufficient statistics.

There are two key points\footnote{These points assume the system is not driven `too hard' so that the probability distribution that characterizes the system does not change during trials or observations. } illuminated by the analogy. The first point
is \emph{the effects of a constraint can be inferred from a subset
$S_s(X)\subset X$ given $\widetilde{\Theta_{S_s(X)}}(S_s(x)|c)$ and
the marginal probability density $\Theta_C(c)$}. And the second is
\emph{the marginal and conditional probability distributions on $C$
belong to the same conjugate family}, i.e.
\begin{equation}
\widetilde{\Theta_C}(c|x)\propto
\Theta_C\left(c\right)\widetilde{\Theta_{S_s(X)}}(S_s(x)|c)\;.
\end{equation}

There is great value in these two key points: We can understand a constrained system through the constraint distribution and a subset of its dynamical variables, the sufficient statistics. Moreover, given a particular
likelihood function and a set of sufficient statistics, the possible
conjugate distributions on $C$ are quite limited. In fact,
consulting a table of conjugate priors for standard distributions,
one can readily find the associated conjugate families.

There are, no doubt, further lessons to be learned about constraints
from the probability correspondence, but at this point we leave the
analogy and return to the CDM scheme of functional integration.

\subsection{Conditional and Conjugate integrators}
Return to $B$ a Banach space of pointed paths, and amend the CDM
scheme with the definition\footnote{The use of $\Theta$ in both the
probability and functional integral context is meant to be
suggestive, but it should be kept in mind that the same symbol is
referring to two distinct objects that should not be confused.}
\begin{definition}\label{def3}
Let $B\equiv X\times Y$ be a Banach product  space, and let each
component Banach space be endowed with CDM scheme data. Define
\begin{equation}
\Theta_{X|Y}(x|y,x'|y'):=\frac{\Theta_{B}(b,b')}{\Theta_{Y}(y,y')}
=\frac{\Theta_{B}(b,b')}{\int_X\Theta_{B}(b,b')
\mathcal{D}_{\Theta_{X},\mathrm{Z}_{X}}x}
\end{equation}
and
\begin{equation}
\mathrm{Z}_{{X'|Y'}}(x'|y'):= \frac{\mathrm{Z}_{B'}(b')}{\mathrm{Z}_{Y'}(y')}
=\frac{\mathrm{Z}_{B'}(b')}{\int_{X'}\mathrm{Z}_{B'}(b')\;d\mu_{X'}(x')}\;.
\end{equation}
These two functionals define an associated conditional integrator by
\begin{equation}\label{integral definition}
  \int_{B}\Theta_{X|Y}(x|y,x'|y')\mathcal{D}_{\Theta_{X|Y},\mathrm{Z}_{X|Y}}x|y
  :=\mathrm{Z}_{{X'|Y'}}(x'|y')\;.
\end{equation}
The space $\mathbf{F}_{X|Y}(B)$ of constrained integrable
functionals consists of functionals defined by
\begin{equation}
\mathrm{F}_{\mu}(x|y):=\int_{B'}\Theta_{X|Y}(x|y,x'|y')\;d\mu(x'|y')
=\int_{B'}\frac{\Theta_{B}(b,b')}{\Theta_{Y}(y,y')}\;d\mu(x'|y')
\end{equation}
where $\mu(x'|y')$ is a conditional measure\footnote{The conditional
measure is well defined as the restriction of $\mu$ to the
appropriate sub-$\sigma$-algebra over $B'$.} on $B'$. Then the linear
integral operator on $\mathbf{F}_{X|Y}(B)$ is given by
\begin{equation}
\int_{B}\mathrm{F}_{\mu}(x|y)\mathcal{D}_{\Theta_{X|Y},\mathrm{Z}_{X|Y}}x|y =\int_{B'}
\mathrm{Z}_{{X'|Y'}}(x'|y')\;d\mu(x'|y')\;.
\end{equation}
\end{definition}

\begin{proposition}\label{prop3.1}
\begin{eqnarray}
\int_{B}\Theta_{X|Y}(x|y,x'|y')\mathcal{D}_{\Theta_{X|Y},\mathrm{Z}_{X|Y}}x|y
&=&\frac{1}{\mathrm{Z}_{Y'}(y')}\int_{B}\Theta_{B}(b,b')
\mathcal{D}_{\Theta_{B},\mathrm{Z}_{B}}b\notag\\ &\big\Downarrow&\notag\\
\mathcal{D}_{\Theta_{X|Y},\mathrm{Z}_{X|Y}}x|y&\;\sim\;&
\frac{\Theta_{Y}(y,\cdot)}{\mathrm{Z}_{Y'}(\cdot)}\mathcal{D}_{\Theta_{B},\mathrm{Z}_{B}}b
\end{eqnarray}
In particular, since the integrator relation holds for any $y'\in
Y'$,
\begin{equation}
\int_{B}\mathrm{F}_{\mu}(x|y)\mathcal{D}_{\Theta_{X|Y},\mathrm{Z}_{X|Y}}x|y
=\frac{1}{\mathrm{Z}_{Y'}(y')}
\int_{B}\mathrm{F}_{\mu}(x|y)\Theta_{Y}(y,y')\mathcal{D}_{\Theta_{B},\mathrm{Z}_{B}}b
\end{equation}
most often with $\langle y',y\rangle= S_s(y)$ or $\langle y',y\rangle=0$ for all $y\in Y$.
\end{proposition}
\emph{Proof.} The proof follows immediately from definition
\ref{def3} and the relevant CDM definitions. $\QED$

Evidently expressing integrals like $\int_B\mathrm{F}(b)\mathcal{D}b$ when
$B$ is a product space requires knowledge of the `marginal'
\emph{and} `conditional' integrators on the component spaces. Of
course, when elements in $X$ and $Y$ are independent, the
conditional integrator on $B$ reduces to a simple product of
standard integrators on $X$ and $Y$. But we anticipate that
constraints induce a dependence between elements in $X$ and $Y$.

Now specialize to the case when $Y$ represents non-dynamical degrees
of freedom --- ostensibly due to constraints. As suggested by the
optical diffraction example, postulate that the physical system is
described by `sufficient statistics'\footnote{In a functional
integral context, `sufficient statistics' is interpreted naturally
as a localization in the space of paths precipitated by some
constraint. For a QM system, constraints restrict the domain of the evolution operator, and it is fruitful to identify `sufficient statistics' with the evolution operator's spectra at $\ti=\ti_a$. In effect, the conditional integrator loosely represents a spectral measure.} and that $\Theta_Y$ and $\Theta_{Y|X}$ belong to the
same conjugate family. Then knowledge of the `likelihood' functional
$\Theta_{X|Y}(x|y,x'|y')$ implies knowledge of the conjugate family
of $\Theta_{Y}(y,y')$ and vice versa. Consequently, the heuristic integral
$\int_B\mathrm{F}(b)\mathcal{D}b$ will be well defined in terms of known
functionals.

Accordingly, the Bayesian analogy suggests the definition:
\begin{definition}
Conjugate integrators are characterized by
\begin{equation}\label{conjugate}
\Theta_{Y|X}(y|x,\cdot)\propto
\Theta_{S_s(X)|Y}(S_s(x)|y,\cdot)\,\Theta_{Y}(y,\cdot)
\end{equation}
where
\begin{equation}
\int_Y\Theta_{Y}(y,y')\mathcal{D}_{\Theta_{Y},\mathrm{Z}_Y}y=\mathrm{Z}_Y(y')
\end{equation}
and the proportionality is fixed by normalization.
\end{definition}
Note that this implies (by Bayes' theorem)
\begin{equation}
\Theta_{X|Y}(x|y,\cdot)\propto\Theta_{B}\left((S_s(y),x),\cdot\right)\;.
\end{equation}
This property suggests the solution strategy
\begin{eqnarray}\label{strategy}
\int_{\widetilde{X}}\mathrm{F}_{\mu}(\widetilde{x})
\mathcal{D}_{\Theta_{\widetilde{X}},\mathrm{Z}_{\widetilde{X}}}\widetilde{x}
&:=&\int_{B}\mathrm{F}_{\mu}(x|y)\mathcal{D}_{\Theta_{X|Y},\mathrm{Z}_{X|Y}}x|y\notag\\
&=:&\int_{B}\widetilde{\mathrm{F}}_{\mu}(S_s(y),x,\cdot)\Theta_Y(y,\cdot)
\mathcal{D}_{\Theta_{B},\mathrm{Z}_{B}}b\notag\\
&=&\int_{X}\left[\int_Y\widetilde{\mathrm{F}}_{\mu}(S_s(y),x,\cdot)
\Theta_Y(y,\cdot)\mathcal{D}_{\Theta_{Y},\mathrm{Z}_{Y}}y\right]
\mathcal{D}_{\Theta_{X},\mathrm{Z}_{X}}x\notag\\
&=:&\int_{X}\widetilde{\mathrm{G}}_{\mu}(x)
\mathcal{D}_{\Theta_{X},\mathrm{Z}_{X}}x
\end{eqnarray}
where the integral on the left is interpreted as a constrained functional
integral, i.e. an integral over the constrained function space
$\widetilde{X}$, the third line employs functional Fubini (Prop. A.3 in \cite{LA2}), and $\widetilde{\mathrm{G}}_{\mu}$ can be interpreted as
a constrained functional depending on the constraints only through sufficient statistics. This is the functional integral analog of
(\ref{Bayes}).

Notice that the statement is equally valid with
$X\leftrightarrow Y$ if one knows some $S_s(X)$; hence suggesting
an alternative solution strategy
\begin{eqnarray}
\int_{\widetilde{X}}\mathrm{F}_{\mu}(\widetilde{x})
\mathcal{D}_{\Theta_{\widetilde{X}},\mathrm{Z}_{\widetilde{X}}}\widetilde{x}
&:=&\int_{B}\mathrm{F}_{\mu}(y|x)\mathcal{D}_{\Theta_{Y|X},\mathrm{Z}_{Y|X}}y|x\notag\\
&=:&\int_{B}\widetilde{\mathrm{F}}_{\mu}(S_s(x),y,\cdot)\Theta_X(x,\cdot)
\mathcal{D}_{\Theta_{B},\mathrm{Z}_{B}}b\notag\\
&=&\int_{Y}\left[\int_X\widetilde{\mathrm{F}}_{\mu}(S_s(x),y,\cdot)\Theta_X(x,\cdot)
\mathcal{D}_{\Theta_{X},\mathrm{Z}_{X}}x\right]
\mathcal{D}_{\Theta_{Y},\mathrm{Z}_{Y}}y\notag\\
&=:&\int_{Y}\widetilde{\mathrm{H}}_{\mu}(y)
\mathcal{D}_{\Theta_{Y},\mathrm{Z}_{Y}}y\;.
\end{eqnarray}
Both strategies can be fruitfully employed depending on one's knowledge of a system's relevant sufficient statistics.

The important point worth emphasizing is that $\Theta_{Y|X}$ and
$\Theta_{Y}$ belong to the same family of integrators, and they are
simply related through the sufficient statistics that describe the
integrator on $X$. This quickly narrows the search for an integrator
associated with a particular constraint.

\section{Conclusion}
We used Bayesian inference theory within the CDM scheme for functional integration to propose a basis for formulating constrained functional integrals. The probability analogy introduces two main ideas. The first idea is that a constrained dynamical system is partially characterized by a subset of its associated function space --- the functional analog of sufficient statistics. (Quite often the subset will be a finite-dimensional subspace isomorphic to some target manifold associated with a physical model.)
The second idea is that a functional integral whose domain is a constrained function space can instead be constructed on an enlarged function space equipped with conditional and conjugate integrators.

However natural the probability analogy may seem, the usefulness of the defined functional integrals rests on their efficacy --- which in turn depends on establishing physically relevant conditional and conjugate integrators. To this end, we describe in detail four particularly pertinent integrator families in Appendix \ref{integrators} and use them in the companion paper \cite{LA3} to re-examine the motivating examples of \S \ref{motivation} in light of our proposed formulation.
\appendix
\section{CDM scheme}\label{appendix 1}
The Cartier/DeWitt-Morette scheme \cite{CA/D-W3}-- \cite{CA/D-M}
defines functional integrals in terms of the data
$(B,\Theta,\mathrm{Z},\mathbf{F}(B))$.

Here $B$ is a separable (usually) infinite dimensional Banach space
with norm $\|\z\|$ where $\z\in B$ is an $L^{2,1}$ map
$\z:[\ti_a,\ti_b]\in\R\rightarrow \C^m$. The dual Banach space $B'\ni
\z'$ is a space of linear forms such that $\zdual_B\in\C$ with an
induced norm given by
\begin{equation}
  \|\z'\|=\sup_{\z\neq 0}|\zdual|/\|\z\|\;.\notag
\end{equation}
Assume $B'$ is separable. Then $B'$ is Polish and consequently
admits complex Borel measures $\mu$.

$\Theta$ and $\mathrm{Z}$ are bounded, $\mu$-integrable functionals
$\Theta:B\times B'\rightarrow \C$ and $\mathrm{Z}:B'\rightarrow\C$. The
functional $\Theta(\z,\cdot)$ can be thought of as the functional
analog of a probability distribution function and $\mathrm{Z}(\z')$ the
associated characteristic functional.

The final datum is the space of integrable functionals
$\mathbf{F}(B)$ consisting of functionals $\mathrm{F}_\mu:B\rightarrow\C$ defined relative
to $\mu$ by
\begin{equation}\label{integrable functions}
\mathrm{F}_{\mu}(\z):=\int_{B'}\Theta(\z,\z')\,d\mu(\z')\;.
\end{equation}
If $\mu\mapsto \mathrm{F}_{\mu}$ is injective, then $\mathbf{F}(B)$ is a
Banach space endowed with a norm $\|\mathrm{F}_{\mu}\|$ defined to be the
total variation of $\mu$.

These data are used to define an integrator
$\mathcal{D}_{\Theta,\mathrm{Z}}\z$ on $B$ by
\begin{equation}\label{integrator}
  \int_{B}\Theta(\z,\z')\,\mathcal{D}_{\Theta,\mathrm{Z}}\z:=\mathrm{Z}(\z')\,.
\end{equation}
This defines an integral operator $\int\limits_B\mathcal{D}_{\Theta,\mathrm{Z}}\z$ on the
Banach space $\mathbf{F}(B)$;
\begin{equation}\label{integral definition}
  \int_{B}\mathrm{F}_{\mu}(\z)\,\mathcal{D}_{\Theta,\mathrm{Z}}\z:=
  \int_{B'}\mathrm{Z}(\z')\,d\mu(\z')=:\int_{B}\mathrm{F}_{\mu}(\z)\,\mathcal{D}_{\Theta,\mathrm{Z}}(\z+b_o)
\end{equation}
for some fixed $b_o\in B$.\footnote{There are ways to motivate translation invariance of the integrator, but here we will simply define it that way.}
The integral operator is a bounded linear form
on $\mathbf{F}(B)$ with
\begin{equation}
\left|\int_B
\mathrm{F}_{\mu}(\z)\,\mathcal{D}_{\Theta,\mathrm{Z}}\z\right|\leq\|\mathrm{F}_{\mu}\|\;.
\end{equation}

\section{Integrator families}\label{integrators}

\subsubsection{Gaussian family}
Before defining Gaussian integrators we establish some terminology. Let $X_a$ be the space of $L^{2,1}$ pointed functions
$x:[\ti_a,\ti_b]\subseteq\R\rightarrow\mathbb{X}$ such that
$x(\ti_a)=:\x_a\in\X$ with $\X$ a real, \emph{flat} differentiable manifold and $\dot{x}(\ti_b)=:\dot{\x}_b\in T_{\x}\X$. The variance $\mathrm{W}:X_a'\times X_a'\rightarrow\C$ is a bilinear form with domain $\mathrm{D}_\mathrm{W}=X_a'$ defined by
\begin{equation}
\mathrm{W}(x_1',x_2'):=\frac{1}{2}\left\{\langle x_1',Gx_2'\rangle +\langle
x_2',Gx_1'\rangle\right\}=:\langle x',G x'\rangle_{\{1,2\}}
\end{equation}
where the covariance $G:X_a'\rightarrow X_a$ is non-negative definite. Associated with the variance is a
\emph{symmetric, closed}\footnote{$\mathrm{Q}$ closed means that its domain $\mathrm{D}_\mathrm{Q}$ can be endowed with a Hilbert space structure. It can be shown that for $\mathrm{Q}$ symmetric and closed there exists a unique self-adjoint operator $A:\mathrm{D}_\mathrm{Q}\rightarrow \mathrm{D}_\mathrm{Q}$ such that $\mathrm{D}_A\subset \mathrm{D}_\mathrm{Q}$ and $\mathrm{Q}(x_1,x_2)=( x_1,Ax_2)$ for any $x_1\in \mathrm{D}_\mathrm{Q}$ and $x_2\in \mathrm{D}_A$ (\cite{BL}, Th. 4.6.8). The boundary form enforces $\mathrm{D}_A= \mathrm{D}_\mathrm{Q}$.} form $\mathrm{Q}:X_a\times X_a\rightarrow\C$;
\begin{equation}
-\mathrm{Q}(x_1,x_2)=\langle Dx,x\rangle_{\{1,2\}}-\mathrm{B}(\xx_1,\xx_2)
\end{equation}
where $D:X_a\rightarrow X'_a$ is a linear map and the mean
path $\bar{x}$ is a critical path determined by $D\bar{x}=0$ and endowed with suitable boundary conditions $\xx(\ti_a)=\x_a$ and
$c_b(\xx(\ti_b),\dot{\xx}(\ti_b),\ldots)=0$. $\mathrm{B}(\xx_1,\xx_2)$ is a symmetric boundary form.\footnote{For example, if $\mathrm{Q}(x_1,x_2)=\int\dot{x}_1\dot{x}_2\,dt$
, then $D=d^2/dt^2$ and $\mathrm{B}(x_1,x_2)=1/2(x_1\dot{x}_2|_{\ti_a}^{\ti_b}+\dot{x}_1x_2|_{\ti_b}^{\ti_a})$.
So $\mathrm{B}(x)=x\dot{x}|_{\ti_a}^{\ti_b}\neq0$ unless $x(\ti_a)=\dot{x}(\ti_b)=0$ or $x(\ti_b)=\dot{x}(\ti_a)=0$. As a less trivial check, the reader can verify that $\mathrm{B}(\bar{x})=-\omega[(\x_a^2+\x_b^2)\cos\omega(\ti_b-\ti_a)-2\x_a\x_b]/\sin\omega(\ti_b-\ti_a)=\mathrm{Q}(x_{cr})$ when $D=d^2/dt^2+\omega^2$ on the space of paths with both end-points fixed.} Note that $\mathrm{Q}(\xx)=\mathrm{B}(\xx)$.

Let $X_{\xx_a}$ be the space $X_a\backslash \mathrm{Ker}(D)$. Then, \emph{restricting to this factor space}, we have $DG=\mathrm{Id}_{X'_{\xx_a}}$ and $GD=\mathrm{Id}_{X_{\xx_a}}$ and so $\mathrm{W}(x')$ and $\mathrm{Q}(x)$ are inverse \emph{up to a boundary form} in this case. Further, any $x\in X_a$ can be reached from a given $\bar{x}$ by $x=\bar{x}+Gx'$ for all $x'\in X'_{{\xx_a}}$. Consequently, each non-trivial zero mode spawns a copy of $X_{\xx_a}$ in $X_a$.\footnote{This brief characterization of $\mathrm{W}$ and $\mathrm{Q}$ can and should be rigorously developed in the context of linear operators on the Hilbert space associated with a constrained function space $\widetilde{X}$. In particular, one should apply results regarding self-adjoint extensions of $D$ and their associated spectra in this context. A thorough study would produce a useful translation dictionary between the rigorous mathematics describing linear operators on Hilbert spaces and their Gaussian functional integral counterparts.}

\begin{definition}\label{Gaussian}
A family of Gaussian integrators $\mathcal{D}\omega_{\xx,\mathrm{Q}}(x)$ is characterized
by\footnote{This definition uses a different normalization from the
usual Gaussian integrator in the CDM scheme. Both definitions are
valid: we choose this normalization because it seems more consistent
with definitions of other integrator families and it highlights the
role of the functional determinant.}
\begin{eqnarray}\label{Gaussian definition}
&&\Theta_{\xx,\mathrm{Q}}(x,x')=e^{2\pi i \langle x',x\rangle-(\pi/\s) \left[\mathrm{Q}(x-\xx)-\mathrm{B}(\xx)\right]}\notag\\
&&\mathrm{Z}_{\xx,\mathrm{W}}(x')=\sqrt{\s}\,\mathrm{Det} (\mathrm{W})^{1/2}e^{2\pi i\langle
x',\xx\rangle-\pi\s \mathrm{W}(x')}
\end{eqnarray}
where $\langle x',x\rangle\in\R$, $\s\in\{1,i\}$, and the functional determinant is
assumed to be well-defined.

The Gaussian integrator family is defined in terms of the primitive integrator $\mathcal{D}x$ by
\begin{equation}
\mathcal{D}\omega_{\xx,\mathrm{Q}}(x):=e^{-(\pi/\s) \left[\mathrm{Q}(x-\xx)-\mathrm{B}(\xx)\right]}\mathcal{D}x=:e^{(\pi/\s)\mathrm{B}(\xx)}\mathcal{D}\omega_{0,\mathrm{Q}}(x)
\end{equation}
where $\mathcal{D}x\equiv\mathcal{D}_{\Theta_{0,\mathrm{Id}}}x$ is characterized by
\begin{equation}
\Theta_{0,\mathrm{Id}}(x,x')=\exp\{2\pi i \langle
x',x\rangle-(\pi/\s)\mathrm{Id}(x)\}\;\;;\;\;
\mathrm{Z}_{0,\mathrm{Id}}(x')=\sqrt{\s}e^{-\pi\s \;\mathrm{Id}(x')}\;.
\end{equation}
\end{definition}

Loosely, the primitive integrator $\mathcal{D}x$ (which is characterized by zero mean and trivial covariance) is
the infinite dimensional analog of the Lebesgue measure on $\R^n$. Note that
$\mathrm{W}$ (and hence $\mathrm{Det}\,\mathrm{W}$), inherits the boundary conditions
imposed on $x$, and note the normalizations
\begin{equation}
\int_{X_0}\mathcal{D}\omega_{0,\mathrm{Id}}(x)=\int_{X_0}e^{-(\pi/\s)\mathrm{Id}(x)}\mathcal{D}x=\sqrt{\s}
\end{equation}
and
\begin{equation}\label{integral def}
\int_{X_a}\mathcal{D}\omega_{\xx,\mathrm{Q}}(x)
=\int\!\!\!\!\!\!\!\!\sum_{\xx}\int_{X_{\xx_a}}\mathcal{D}\omega_{\xx,\mathrm{Q}}(x)=\int\!\!\!\!\!\!\!\!\sum_{\xx} \sqrt{\s}\,\mathrm{Det}
 (\mathrm{W})^{1/2}e^{(\pi/\s) \mathrm{B}(\xx)}\;.
\end{equation}
Three points to emphasize: The fiducial Gaussian integrator $\mathcal{D}\omega_{0,\mathrm{Id}}(x)$ is associated with the bona fide Banach space $X_0=X_{\xx_0}$ where the primitive integrator is translation invariant, i.e. $\mathcal{D}(x_1-x_2)=\mathcal{D}(x_1)$. For any given $\xx$, the middle integral in (\ref{integral def}) can therefore be written as an integral over $X_0$ by a change of integration variable $x-\xx\mapsto\widetilde{x}$ with $\widetilde{x}(\ti_a)=0$ since the primitive integrator is translation invariant. Finally, since there is a copy of $X_{\xx_a}$ for each non-trivial zero mode, we see clearly why an integral over the full space $X_a$ must include a sum/integral over all $\xx$.

The resemblance between the functional form of $\mathrm{Z}(x')$ and the exponential multiplying the primitive integrator motivates the standard practice in quantum field theory of
defining the effective action functional. First, note that
\begin{equation}
e^{(\pi/\s)\Gamma'_{\xx}(x')}:=e^{2\pi i\langle x',\xx\rangle-\pi\s \mathrm{W}(x')}
\end{equation}
is nothing other than the characteristic functional of the Gaussian integrator $\mathcal{D}\omega_{\xx,\mathrm{Q}}(x)$. Moreover,
\begin{eqnarray}
\frac{1}{2\pi i}\left.\frac{\delta}{\delta x'(\ti)}\frac{1}{\mathrm{Z}_{\xx,\mathrm{W}}(0)}e^{-(\pi/\s)\Gamma'_{\xx}(x')}\right|_{x'=0}
 &=&\left.\left(\xx(\ti)-\frac{\s}{2i}\frac{\delta \mathrm{W}(x')}{\delta x'(\ti)}\right)\right|_{x'=0}\notag\\
 &=&\xx(\ti)\notag\\
&=&\frac{1}{\mathrm{Z}_{\xx,\mathrm{W}}(0)}\int_{X_{\xx_a}}x(\ti)\;\mathcal{D}\omega_{\xx,\mathrm{Q}}(x)\notag\\
&=:& E(x)(\ti)\;.
\end{eqnarray}

So define the effective action evaluated at $E(x)$ by
\begin{equation}\label{effective action}
\Gamma_{\widetilde{E}(x')}(E(x)):=\Gamma'_{E(x)}(\widetilde{E}(x'))
\end{equation}
where, for a given $x'$, the dual expectation $\widetilde{E}(x')$ is determined by
\begin{equation}
\langle\widetilde{E}(x'),x\rangle=\langle x',E(x)\rangle\;\;\forall x\in X_a\;.
\end{equation}

Essentially, the weighted sum over all zero modes of the exponentiated effective action is the expectation of $e^{-[\mathrm{Q}-\mathrm{B}]}$ with respect to the primitive integrator $\mathcal{D} x$. More precisely,
\begin{equation}
\int\!\!\!\!\!\!\!\!\sum_{\xx}\sqrt{\s}\,\mathrm{Det}
 (\mathrm{W})^{1/2}e^{(\pi/\s)\Gamma_{\widetilde{E}(x')}(E(x))}=\int_{X_a}e^{-(\pi/\s)\left[\mathrm{Q}(x-\bar{x})-\mathrm{B}(\bar{x})\right]}\;\mathcal{D}x\;.
\end{equation}
Notice that, since $\mathrm{Q}$ is quadratic,  the functional integral is easily evaluated once $\xx$ is known and the effective action is trivially $\Gamma_{\widetilde{E}(x')}(E(x))=\mathrm{B}(\xx)=\mathrm{Q}(\xx)$. However, Gaussian integrators can be readily generalized to non-Gaussian integrators based on non-quadratic action functionals $\mathrm{S}:X_a\times X_a\rightarrow\C$ in the CDM scheme --- in which case the effective action becomes a useful tool.

To see how conditional Gaussian integrators work, form the product
space $X_a\times Y_a$. Suppose a Gaussian integrator on $X_a\times
Y_a$ is characterized by a positive definite quadratic form
$\widetilde{\mathrm{Q}}$ with mean $\bar{m}$ and \emph{vanishing boundary
term}. Put $\bar{m}=(\xx,\bar{y})$ and
\begin{equation}\label{product space G}
\widetilde{G}=\left(
                \begin{array}{cc}
                  G_{xx} &  G_{xy} \\
                   G_{yx} &  G_{yy} \\
                \end{array}
              \right)\;.
\end{equation}
Then\footnote{It can be shown that $\mathrm{Q}_X$ and $\mathrm{Q}_Y$ are positive
definite since $\widetilde{G}$ is positive definite.}
\begin{equation}
\widetilde{\mathrm{Q}}\left((x,y)-\bar{m}\right)=\mathrm{Q}_X(x-\bar{m}_{x|y})+\mathrm{Q}_Y(y-\bar{y})
\end{equation}
where $\mathrm{Q}_Y(y_1,y_2)=\left\langle D_{yy}y_1,y_2\right\rangle$,
\begin{equation}
\mathrm{Q}_X(x_1,x_2)=\langle
\left(G_{xx}-G_{xy}D_{yy}G_{yx}\right)^{-1}x_1,x_2\rangle
\end{equation}
and
\begin{equation}
\bar{m}_{x|y}=\xx+G_{xy}D_{yy}(y-\bar{y})\;.
\end{equation}

 So the Gaussian integrator on $X_a\times Y_a$ is
\begin{equation}
\mathcal{D}\omega_{\widetilde{m},\widetilde{\mathrm{Q}}}(x,y) :=e^{-(\pi/\s)
\widetilde{\mathrm{Q}}((x,y)-\bar{m})}\mathcal{D}(x,y)\;.
\end{equation}
On the other hand,
\begin{equation}
\mathcal{D}\omega_{\bar{y},\mathrm{Q}_Y}(y) :=e^{-(\pi/\s)
\mathrm{Q}_Y(y-\bar{y})}\mathcal{D}y\;.
\end{equation}
Therefore, the \emph{conditional Gaussian integrator} is
\begin{equation}\label{conditional integrator}
\mathcal{D}\omega_{\bar{m}_{x|y},\mathrm{Q}_{X|Y}}(x|y) := e^{-(\pi/\s)
\mathrm{Q}_{X}(x-\bar{m}_{x|y})}\mathcal{D}(x,y)
\end{equation}
which yields
\begin{equation}\label{conditional integral}
\int_{\mathrm{B}_a}\mathcal{D}\omega_{\bar{m}_{x|y},\mathrm{Q}_{X|Y}}(x|y) =\sum_{\bar{m}_{x|y}}
\frac{\mathrm{Det}(\mathrm{Q}_X+\mathrm{Q}_Y)^{-1/2}}{\mathrm{Det}(\mathrm{Q}_Y)^{-1/2}}\;.
\end{equation}

In particular, let $M:X_0\rightarrow Y_0$ be a homeomorphism such
that $\mathrm{Q}_1=\mathrm{Q}_2\circ M$. If $Y_0=X_0$ then $G_{xy}=G_{yx}=0$ since the
$x$ are independent Gaussian variables. Then
formally,
\begin{equation}
\mathcal{D}\omega_{0,\mathrm{Q}_1}(x)
=\frac{\mathcal{D}\omega_{0,\mathrm{Q}_1}(x|y)}
{\mathcal{D}\omega_{0,\mathrm{Q}_2}(y|x)}\mathcal{D}\omega_{0,\mathrm{Q}_2}(y)\;.
\end{equation}
But
\begin{equation}
\frac{\mathcal{D}\omega_{0,\mathrm{Q}_1}(x|y)}
{\mathcal{D}\omega_{0,\mathrm{Q}_2}(y|x)}\sim\frac{\mathrm{Det}(\mathrm{Q}_1)^{-1/2}}{\mathrm{Det}(\mathrm{Q}_2)^{-1/2}}
\end{equation}
so we get the standard result for a change of covariance;
\begin{equation}\label{covariance change}
\int_{X_0}e^{-(\pi/\s) \mathrm{Q}_2(x)}\mathcal{D}_1x
=\frac{\mathrm{Det}(\mathrm{Q}_2)^{-1/2}}{\mathrm{Det}(\mathrm{Q}_1)^{-1/2}}
\end{equation}
where $\mathcal{D}_1x$ is the primitive  integrator on $X_0$.
Obviously the same condition holds for $1\leftrightarrow 2$ with
$\mathcal{D}_2x$ the primitive integrator on $M(X_0)$.

\subsubsection{Complex Gaussian family}
The previous subsection took the parameter
$\s\in\{1,i\}$.\footnote{That Gaussian integrators based on
non-negative definite \emph{real} $G$ can be defined for
$\s\in\{1,i\}$ reflects the validity of the
Schr\"{o}dinger$\leftrightarrow$diffusion correspondence through
analytic continuation. However, analytic continuation does not
maintain this correspondence in general. It is natural to conjecture
that the analytic continuation
Schr\"{o}dinger$\leftrightarrow$diffusion correspondence will break
down precisely when $G_{zz}$ and/or $G_{\underline{zz}}$, defined
below, do not vanish.} This restriction can be lifted by defining a
complex Gaussian integrator.

\begin{definition} Let $\mathrm{Z}_{a}^2$ be the space of $L^{2,1}$ pointed
functions
$(z,\underline{z}):[\ti_a,\ti_b]\subseteq\R\rightarrow\M^\C$ such
that $(z,\underline{z})(\ti_a)=:(z_a,\underline{z}_a)\in\M^\C$ with
$\M^\C$ a flat complex manifold. A complex Gaussian family of integrators
$\mathcal{D}\omega_{\w,\mathrm{Q}^\C}(w)$ on $\mathrm{W}_a\equiv \mathrm{Z}_{a}^2$ is
characterized by
\begin{eqnarray}
&&\Theta_{\w,\mathrm{Q}^\C}(w,w')=e^{2\pi i \langle w',w\rangle-\pi \left[\mathrm{Q}^\C(w-\w)-\mathrm{B}^\C(\w)\right]}\notag\\
&&\mathrm{Z}_{\w,\mathrm{W}^\C}(w')=\mathrm{Det}(\mathrm{W}{^\C})^{1/2} e^{2\pi i \langle
w',\w\rangle-\pi \mathrm{W}^\C(w')}
\end{eqnarray}
where $w:=(z,\underline{z})\in \mathrm{W}_a$, $w'=(z',{{\underline{z}'}})\in
\mathrm{W}'_a$, and $\langle w',w\rangle\in\C$. The complexified variance
$\mathrm{W}^\C(w_1',w_2')=\langle w_1',G^\C w_2'\rangle$ where the complex
covariance matrix $G^\C$ has the block form
\begin{equation}
G^\C=\left(
       \begin{array}{cc}
         G_{\underline{z}\,z} & G_{\underline{z}\underline{z}} \\
         G_{zz} & G_{z\,\underline{z}} \\
       \end{array}
     \right)
\end{equation}
with $\Re(\langle w_1',G^\C w_2'\rangle)\geq 0$ and $G^\C$ not
necessarily Hermitian.\footnote{If $\underline{z}=z^\ast$ then
$(G^\C)^\dag=G^\C$.} As in the real case, put
\begin{equation}
\mathcal{D}\omega_{\w,\mathrm{Q}^\C}(w)=e^{-\pi \left[\mathrm{Q}^\C(w-\w)-\mathrm{B}^\C(\w)\right]}\mathcal{D}w
\end{equation}
where $\mathcal{D}w$ is characterized by
\begin{equation}
\Theta_{0,\mathrm{Id}}(w,w')=\exp\{2\pi i \langle
w',w\rangle-\pi\mathrm{Id}(w)\}\;\;;\;\; \mathrm{Z}_{0,\mathrm{Id}}(w')=e^{-\pi
\mathrm{Id}(w')}\;.
\end{equation}
\end{definition}

At the level of functional integrals, evidently there is  little
difference between the real and complex Gaussian families. The value
in the complex case comes when the domain of integration is
localized yielding complex line integrals.

\subsubsection{Gamma family}
\begin{definition}
Let $\Ta$ be the space of continuous pointed maps
$\tau:(\mathbb{T}_+,\ti_a)\rightarrow(\C_+,1)$ where $\mathbb{T}_+\subseteq\R_+$ and $\C_+:=\R_+\times i\R$ is the right-half
complex plane. $T_0$ is an abelian group under point-wise multiplication in the first component and point-wise addition in the second. Let $\beta'$ be a fixed element in the dual group $\T$ of linear characters
$\tau':\Ta\rightarrow\C$. A gamma family of integrators
$\mathcal{D}\gamma_{\alpha,\beta'}(\tau)$ on ${T_0}$ is characterized
by\footnote{This definition is somewhat modified from the original
definition in \cite{LA2}: the old definition was in terms of a particular realization of $\beta'$ evaluated on a subspace of $T_0$ rendering it an ordinary function.}
\begin{eqnarray}
&&\Theta_{\alpha,\beta'}(\tau,\tau')= e^{ i\dual-\langle\beta',\tau\rangle}\,\tau^\alpha \notag\\
&&\mathrm{Z}_{\alpha,\beta'}(\tau')
=\mathrm{Det}({\beta'}-i{\tau'})^{-\alpha}
\end{eqnarray}
where $\alpha\in\C$, $\tau^\alpha$ is defined point-wise by
$\tau^\alpha(\ti):=e^{\alpha\log\tau(\ti)}$ with the principal value prescription\footnote{So the branch cut lies outside the range of $\tau$.} for $\log\tau(\ti)$,
and the functional determinant $\mathrm{Det}({\beta'}-i{\tau'})$ is assumed to be well-defined.

The integrator family is defined in terms of the primitive
integrator $\mathcal{D}\tau$ by
\begin{equation}\label{gamma}
\mathcal{D}\gamma_{\alpha,\beta'}(\tau) :=
e^{-\langle\beta',\tau\rangle}\tau^{\alpha}\, \mathcal{D}\tau
\end{equation}
where $\mathcal{D}\tau$ is
characterized by
\begin{equation}
\Theta_{0, Id'}(\tau,\tau')=\exp\{ i
\dual-\langle Id',\tau\rangle\}\;\;;\;\; \mathrm{Z}_{0, Id'}(\tau')=\Gamma(0)\;.
\end{equation}
\end{definition}

Whereas the primitive integrator $\mathcal{D}x$ is the infinite dimensional
analog of the translation invariant measure on $\R^n$; the
 primitive integrator $\mathcal{D}\tau$, for real $\tau(\ti)$, is the analog of the scale invariant
measure on $\R_+$. Note that (\ref{gamma}) requires $|\langle\beta',\tau\rangle|\geq0$ to be well-defined.

 Experience indicates that a prominent sufficient statistic characterizing gamma-type paths is an upper bound  $|\tau(\ti)|\leq|c|$ for all $\ti\in[\ti_a,\ti_b]$ and for some constant $c\in\C_+$ --- much like fixed end-points can characterize the sufficient statistics associated with gaussian paths. The obvious tool to enforce this constraint is the functional analog of Heaviside; yielding a `cut-off' gamma family that generalizes the previous definition but reduces to it as $|c|\rightarrow\infty$.

\begin{definition}
Let $\Ta$ be the space of continuous pointed maps
$\tau:(\mathbb{T}_+,\ti_a)\rightarrow(\C_+,1)$. Let $\beta'$ be a fixed element in the dual group $\T$  and fix some fiducial $\tau_o\in\Ta$ such that $\langle\beta',\tau_o\rangle=c\in\C_+$. A lower gamma family of integrators
$\mathcal{D}\gamma_{\alpha,\beta',c}(\tau)$ on $T_0$ is characterized
by
\begin{eqnarray}
&&\Theta_{\alpha,\beta'}(\tau,\tau')= e^{ i\dual-\langle\beta',\tau\rangle}\,\tau^\alpha \notag\\
&&\mathrm{Z}_{\alpha,\beta',c}(\tau')
=\frac{\gamma\left(\alpha,c\right)}
{\mathrm{Det}(\beta'-i{\tau'})^\alpha}
\end{eqnarray}
where  $\gamma\left(\alpha,c\right)$ is
the lower incomplete gamma functional given by
\begin{equation}
\gamma\left(\alpha,c\right)
=\Gamma(\alpha)e^{-c}\sum_{n=0}^\infty
\frac{(c)^{\alpha+n}}{\Gamma(\alpha+n+1)}\;,
\end{equation}
and the functional determinant $\mathrm{Det}(\beta'-i{\tau'})$ is assumed to be well-defined.

An upper gamma family of integrators
$\mathcal{D}\Gamma_{\alpha,\beta',c}(\tau)$ is defined
similarly where
\begin{equation}
\Gamma\left(\alpha,c\right)
=\Gamma(\alpha)-\gamma\left(\alpha,c\right)
\end{equation}
is the upper incomplete gamma functional.
\end{definition}

Using this notion, the fiducial gamma integrator represented by
$\mathcal{D}\gamma_{0, Id',\infty}(\tau)$ (equivalently
$\mathcal{D}\Gamma_{0, Id',0}(\tau)$). It is
normalized up to a factor of $\Gamma(0);$
\begin{equation}
\frac{1}{\Gamma(0)}\int_{{\Ta}}\mathcal{D}\gamma_{0, Id',\infty}(\tau)=1
=\frac{1}{\Gamma(0)}\int_{{\Ta}}\mathcal{D}\Gamma_{0, Id',0}(\tau)\;,
\end{equation}
but the other family members yield
\begin{equation}\label{gamma normalization}
\frac{1}{\Gamma(\alpha)}\int_{{\Ta}}\mathcal{D}\gamma_{\alpha,\beta',\infty}(\tau)
=\mathrm{Det}\beta'^{-\alpha}
 =\frac{1}{\Gamma(\alpha)}\int_{{\Ta}}\mathcal{D}\Gamma_{\alpha,\beta',0}(\tau)\;.
\end{equation}
Eventually in applications we will run into factors of $\int_{T_0}\mathcal{D}\tau$. Rather than normalizing everything by constantly dividing out this factor, we will \emph{define} it to be $\int_{T_0}\,\mathcal{D}\tau=1$. This is consistent with $\lim_{z\rightarrow0}1/{z^0}=1$ formally applied to (\ref{gamma normalization}).

Put ${B}_a=X_a\times {\Ta}$. For $\Theta_X$ Gaussian and $\Theta_T$
gamma, use the relation for conjugate integrators to get
\begin{eqnarray}
\int_{{B}_a}\Theta_{X|T}(x|\tau,\cdot)\mathcal{D}_{\Theta_{X|T},\mathrm{Z}_{X|T}}x|\tau
&=&\int_{{B}_a}\Theta_{X|T}(x|\tau,\cdot)\frac{\Theta_{T}(\tau,\cdot)}{\mathrm{Z}_{T'}}
\mathcal{D}_{\Theta_{\mathrm{B}},\mathrm{Z}_{\mathrm{B}}}b\notag\\
&\propto&\int_{{B}_a}\Theta_{S_s(T)|X}(S_s(\tau)|x,\cdot)
\frac{\Theta_X(x,\cdot)\Theta_{T}(\tau,\cdot)}{\mathrm{Z}_{T'}}
\mathcal{D}_{\Theta_{\mathrm{B}},\mathrm{Z}_{\mathrm{B}}}b\;.\notag\\
\end{eqnarray}
This suggests that integrals of conditional functionals on
$X_a\times {\Ta}$ be understood as
\begin{equation}
\int_{{B}_a}\mathrm{F}_{\mu}(x|\tau)\mathcal{D}_{\Theta_{X|T},\mathrm{Z}_{X|T}}x|\tau\;
=\int_{{B}_a}\widetilde{\mathrm{F}}_{\mu}(S_s(\tau),x)\mathcal{D}\omega_{\xx,\mathrm{Q}}(x)
\mathcal{D}\gamma_{\alpha,\beta',c}(\tau)
\end{equation}
when $S_s(\tau)$ is a sufficient statistic for the integrator family
characterized by $\Theta_X$. This is just a specialization of the
solution strategy (\ref{strategy}), and it plays a prominent role in the solution of differential equations defined on bounded regions.

\subsubsection{Poisson family}\label{appendix 2}
Take the lower gamma integrator and regularize by replacing
$\gamma(\alpha,c)$ with the regularized lower
incomplete gamma function
$P(\alpha,c):=\gamma(\alpha,c)/\Gamma(\alpha)$.
Restrict the parameters such that $\alpha=n\in\mathbb{N}$, $\beta'= \lambda Id'$,
and $\mathrm{Det}(\beta')=\lambda\mathrm{Det}(Id'):=\lambda$.

Note that, for $N\in
Pois(c)$ a Poisson random variable, we have
\begin{equation}
Pr(N<n)=\sum_{k<n}e^{-c}\frac{(c)^k}{k!}\;.
\end{equation}
Hence,
\begin{equation}
Pr(N\geq n)=\sum_{k=n}^\infty
e^{-c}\frac{(c)^k}{k!}=P(n,c)
=\frac{1}{\Gamma(n)}\int_{{\Ta}}
\mathcal{D}\gamma_{n, Id',c}(\tau)
\end{equation}
which, in particular, implies
\begin{equation}
\frac{1}{\Gamma(0)}\int_{{\Ta}}\;\mathcal{D}\gamma_{0, Id',c}(\tau)
=\sum_{k=0}^\infty e^{-c}\frac{(c)^k}{k!}\;.
\end{equation}
On the other hand,
\begin{equation}
e^{-c}\frac{(c)^k}{k!}=\frac{e^{-c}}{k!}
\int_0^{c}\cdots\int_0^{c} \;d\tau_1,\ldots,d\tau_k\;.
\end{equation}

Not surprisingly, $Pois(c)$ is closely related to the
lower gamma integrator which motivates the following
definition:
\begin{definition}\label{dirac}
Let $\Ta$ be the space of continuous pointed maps
$\tau:(\mathbb{T}_+,\ti_a)\rightarrow(\C_+,1)$
endowed with a lower gamma family of integrators. Let $\alpha=n\in\mathbb{N}$ and
$\langle \beta',\tau_o\rangle=c$
with $c\in\C_+$. The Poisson integrator family
$\mathcal{D}\pi_{n,\beta',c}(\tau)$ is characterized by
\begin{eqnarray}
&&\Theta_{n,\beta'}(\tau,\tau')=e^{i\dual-\langle \beta',\tau\rangle} \tau^n\notag\\
&&\mathrm{Z}_{n, \beta',c}(\tau') = \frac{P\left(n, c\right)}
{\mathrm{Det}\left( \beta'-i{\tau'}\right)^n}\;.
\end{eqnarray}
The Poisson family is defined in terms of the primitive
integrator $\mathcal{D}\tau$ by
\begin{equation}
\mathcal{D}\pi_{n,\beta'}(\tau) :=
e^{-\langle \beta',\tau\rangle}\tau^n\,
\mathcal{D}\tau\;.
\end{equation}
\end{definition}
Note the normalization of the fiducial Poisson integrator
\begin{equation}
\int_{{\Ta}}\;\mathcal{D}\pi_{0,\beta',c}(\tau) =1\;,
\end{equation}
and the rest of the family
\begin{equation}
\int_{{\Ta}}\;\mathcal{D}\pi_{n,\beta',c}(\tau)
=P(n, c)\;.
\end{equation}

For quantum physics applications, $\Re(\tau(\ti))=0$ so that $\tau:(\mathbb{T}_+,\ti_a)\rightarrow(i\R,0)$. In this restricted case $T_0$ becomes a Banach space over $\R$, and  it is useful to define the `shifted' Poisson integrator by
\begin{equation}
\mathcal{D}\widehat{\pi}_{n,\beta',\tau_o}(\tau) :=
e^{-\langle \beta',(\tau-\tau_0)\rangle}\tau^n\,
\mathcal{D}\tau\;.
\end{equation}
Then use the shifted Poisson integrator to define the Poisson expectation of $\beta'$ with respect to $\tau_0$;
\begin{equation}
{\langle{\beta'}\rangle}_{\tau_o}:=\int_{{\Ta}}
\;\mathcal{D}\widehat{\pi}_{0,\beta',\tau_o}(\tau)=e^{\langle \beta',\tau_0\rangle}\;.
\end{equation}
If we take $\langle \beta',\tau_0\rangle=i\int_{\ti_a}^{\ti_b}\beta(\ti)\;d\ti$, then
\begin{eqnarray}
\langle \beta'\rangle_{\tau_o}=
\sum_{n=0}^\infty \frac{i^n}{n!}
\int_{\ti_a}^{\ti_b}\beta'(\ti_1)\cdots\int_{\ti_a}^{\ti_b}\beta'(\ti_n) \;d\ti_1,\ldots,d\ti_n\notag\\
=\sum_{n=0}^\infty
\int_{\ti_a}^{\ti_b}i\beta'(\ti_1)\int_{\ti_1}^{\ti_b}i\beta'(\ti_2)\cdots\int_{\ti_n}^{\ti_b}i\beta'(\ti_n) \;d\ti_1,\ldots,d\ti_n
\end{eqnarray}
where $\ti_a\leq\ti_1<\cdots<\ti_n\leq\ti_b$. Note that $\frac{\partial}{\partial\ti_b}\langle \beta'\rangle_{\tau_o}=i\beta'(\ti_b)\langle \beta'\rangle_{\tau_o}$, so the Poisson expectation solves a first-order evolution equation.

\end{document}